\documentclass[iop]{emulateapj}

\usepackage{amsmath}
\usepackage{graphicx}
\usepackage{natbib}
\usepackage{color}

\shorttitle{}
\shortauthors{Park et al.}

\newcommand{\Hb}{\rm H{$\beta$}}
\newcommand{\OIII}{[O {\small III}]}
\newcommand{\FeII}{Fe {\small II}}
\newcommand{\HeII}{He {\small II}}

\newcommand{\mbh}{M$_{\rm BH}$}     
\newcommand{\ergs}{erg~s$^{\rm -1}$}
\newcommand{\msun}{M$_{\odot}$}       
\newcommand{\msigma}{M$_{\rm BH}-\sigma_{*}$}
\newcommand{\kms}{km~s$^{\rm -1}$}
\newcommand{\sigmastar}{$\sigma_{*}$}
\newcommand{\lam}{$\lambda$}

\newcommand{\lagcen}{$\tau_{\rm cen}$}
\newcommand{\lagpeak}{$\tau_{\rm peak}$}

\newcommand{\sigmaline}{$\sigma_{\rm line}$}

\newcommand{\fwhb}{\ensuremath{\mathrm{FWHM}_\mathrm{H{\beta}}}}

\newcommand{\lf}{\ensuremath{L_{5100}}}

\begin{document}
\title{Reverberation Mapping of PG 0934+013 with the Southern African Large Telescope}
\author{Songyoun Park$^{1}$}
\author{Jong-Hak Woo$^{1}$}
\author{Encarni Romero-Colmenero$^{2,3}$}
\author{Steven M. Crawford$^{2}$}
\author{Dawoo Park$^{1}$}
\author{Hojin Cho$^{1}$}
\author{Yiseul Jeon$^{1}$}
\author{Changsu Choi$^{1}$}
\author{Aaron J. Barth$^{4}$}
\author{Liuyi Pei$^{4}$}
\author{Ryan C. Hickox$^{5}$}
\author{Hyun-Il Sung$^{6}$}
\author{Myungshin Im$^{7}$}

\affil{
$^1$Department of Physics and Astronomy, Seoul National University, Seoul 08826, Republic of Korea; woo@astro.snu.ac.kr \\
$^2$South African Astronomical Observatory, P.O. Box 9, Observatory 7935, Cape Town, South Africa \\
$^3$Southern African Large Telescope Foundation, P.O. Box 9, Observatory 7935, Cape Town, South Africa \\
$^4$Department of Physics and Astronomy, 4129 Frederick Reines Hall, University of California, Irvine, CA 92697-4575, USA \\
$^5$Department of Physics \& Astronomy, Dartmouth College, Hanover, NH 03755, USA \\
$^6$Korea Astronomy and Space Science Institute, Daejeon 34055, Republic of Korea \\
$^7$CEOU, Astronomy Program, Department of Physics \& Astronomy, Seoul National University, Seoul 08826, Republic of Korea \\
}

\begin{abstract}
We present the variability and time lag measurements of PG 0934+013 based on a photometric and spectroscopic monitoring campaign
over a two year period.
We obtained 46 epochs of data from the spectroscopic campaign, which was carried out using the 
Southern African Large Telescope with $\sim$1 week cadence over two sets of 4 month-long observing period,
while we obtained 80 epochs of \textit{B}-band imaging data using a few 1-m class telescopes. 
Due to the seven month gap between the two observing periods, we separately measured the time lags of broad emission lines 
including H$\beta$, by comparing
the emission line light curve with the \textit{B}-band continuum light curve using the cross-correlation function techniques. 
We determined the H$\beta$ lag, $\tau_{\rm cent} = 8.46^{+2.08}_{-2.14}$ days in the observed-frame based on Year 2 data, while the time lag from Year 1 data was not reliably determined. Using the rms spectrum of Year 2 data, we measured the \Hb\ line dispersion \sigmaline = 668 $\pm$ 44 \kms\ after correcting for the spectral resolution. Adopting a virial factor f = 4.47 from \citet{Woo2015}, we determined the black hole mass \mbh\ = $3.13 ^{+0.91} _{-0.93} \times 10^{6}$ \msun based on the \Hb\ time lag and velocity. 

\end{abstract}

\section{INTRODUCTION} \label{section:intro}

The observed scaling relations between black hole mass (M$_{\rm BH}$) and the properties of inactive and active galaxies 
provide interesting constraints in understanding galaxy formation and evolution \citep[see][]{KormendyHo2013}. 
To investigate the role of BHs in the context of BH-galaxy coevolution \citep[e.g.,][]{Peng2006,Woo2006,Treu2007,Jahnke2009,Merloni2010,Bennert2010,Bennert2011b,Cisternas2011,Bennert2015,Park2015} as well as the physics of BH accretion and related phenomena \citep[e.g.,][]{WooUrry2002, Kollmeier2006,Davis2007,Bentz2013,Woo+16},
BH masses need to be accurately determined. 

For active galactic nuclei (AGNs),  M$_{\rm BH}$ can be determined with the reverberation mapping technique \citep{Blandford1982,Peterson1993},
assuming that the kinematics of the broad-line region (BLR) gas are governed by the gravitational potential of the central BH. 
The presence of broad emission lines and flux variability, as one of the main characteristics of type 1 AGNs, allows us to probe the geometry, structure, and kinematics of BLR. The light-travel time to the BLR can be determined by measuring the time delay of variations between continuum and emission lines, which is then translated to the BLR size. The measured BLR size using \Hb\ emission line and AGN continuum luminosity at 5100\AA\ has been reported in the literature, ranging from a few light days to a couple of hundred light days in the rest-frame \citep[e.g.,][]{Bentz2013}. As a measure of the gas velocity, either the second moment (line dispersion; \sigmaline) or the full-width half maximum (FWHM) of the broad emission line profile is typically adopted from the mean or rms spectra generated based on multi-epoch data \citep[see e.g.,][]{Peterson2004, Park2012}.
Combining the time lag and the width of broad emission lines, mass can be determined as \mbh\ $\propto R_{\rm BLR}~ V^2$, where $R_{\rm BLR}$ is the measured time lag between AGN continuum and the BLR, while $V$ is velocity measured from the width of the broad-emission line \citep{Peterson1993}.

Reverberation mapping data are of paramount importance since they provide the fundamental calibration for indirect \mbh\
estimators. Consequently, \mbh\ of distant type 1 AGNs can be estimated based on a single-epoch spectrum, 
instead of a directly measured time lag from a long-term monitoring campaign \citep{Kaspi+00, Vestergaard2006, McGill+08, Woo2015, Park+17}. 
In this method AGN continuum luminosities are used as a proxy for $R_{\rm BLR}$ based on the empirical size-luminosity relation between the measured R$_{\rm BLR}$ and AGN luminosity at 5100 \AA\ \citep[e.g.,][]{Bentz2013}.  
However, these indirect mass estimates are much more uncertain than reverberation masses
\citep[a factor of 2-3 in the case of \Hb-based mass estimates;][]{Woo2010, Park2012} and mainly rely on the limited calibration based on 
the relatively small sample of reverberation-mapped AGNs \citep[e.g.,][]{Woo+13,Grier2013,Woo2015}. 
Thus, enlarging the reverberation-mapped AGNs can provide better calibration and reduce the systematic uncertainties in
the single-epoch mass estimators. 

The reverberation mapping technique has been successfully applied to several dozens of AGNs to date 
\citep[e.g.,][]{Wandel1999, Kaspi+00,Peterson2004,Bentz2009b,Barth2011,Grier2013, Barth2015,Fausnaugh+17}. 
Various reverberation projects, including campaigns utilizing multi-object spectrographs or robotic monitoring,
are currently devised or underway \citep[e.g.][]{Valenti2015, Shen2016}.
One of the difficulties of the reverberation mapping is that a long-term spectroscopic
monitoring with relatively short cadence is required to obtain accurate time
lag measurements. So far, most reverberation studies have been performed with relatively small
telescopes while it would be more efficient with larger-aperture and
service mode telescopes \citep[e.g., see a recent reverberation study with
the HET by][]{Rafter2011}.

As a pilot study with the Southern African Large Telescope (SALT), which provides
the unique capability of quasi-daily access, efficient service observations, and large photon-collecting power,
we performed a reverberation mapping campaign of a nearby Seyfert 1 galaxy PG 0934+013 
at R.A. = 09:37:1.0, Dec. = $+$01:05:43, and redshift z = 0.0503 \citep{Veroncetty2001}.
The target is also known as Mrk 707, which was classified as a narrow-line Seyfert 1 galaxy since the FWHM of the \Hb\ line
was reported to be less than 2000 \kms\ \citep[e.g., 1320 \kms\ by][]{BG92}. The host galaxy is a spiral galaxy with clear signs of spiral arms and  a bar structure
(see S 4.2). We selected this target among bright AGNs
from the Palomar Bright Quasar Survey \citep{Schmidt&Green83}, that are observable at the SALT at least over 4 months per year.
In this paper, we present the results from our reverberation mapping campaign, which was carried out over two seasons, using the SALT for spectroscopy and three smaller telescopes for photometry.

\section{Observations and Data Reduction}

\begin{figure}
	\centering
	\includegraphics[width=0.42\textwidth]{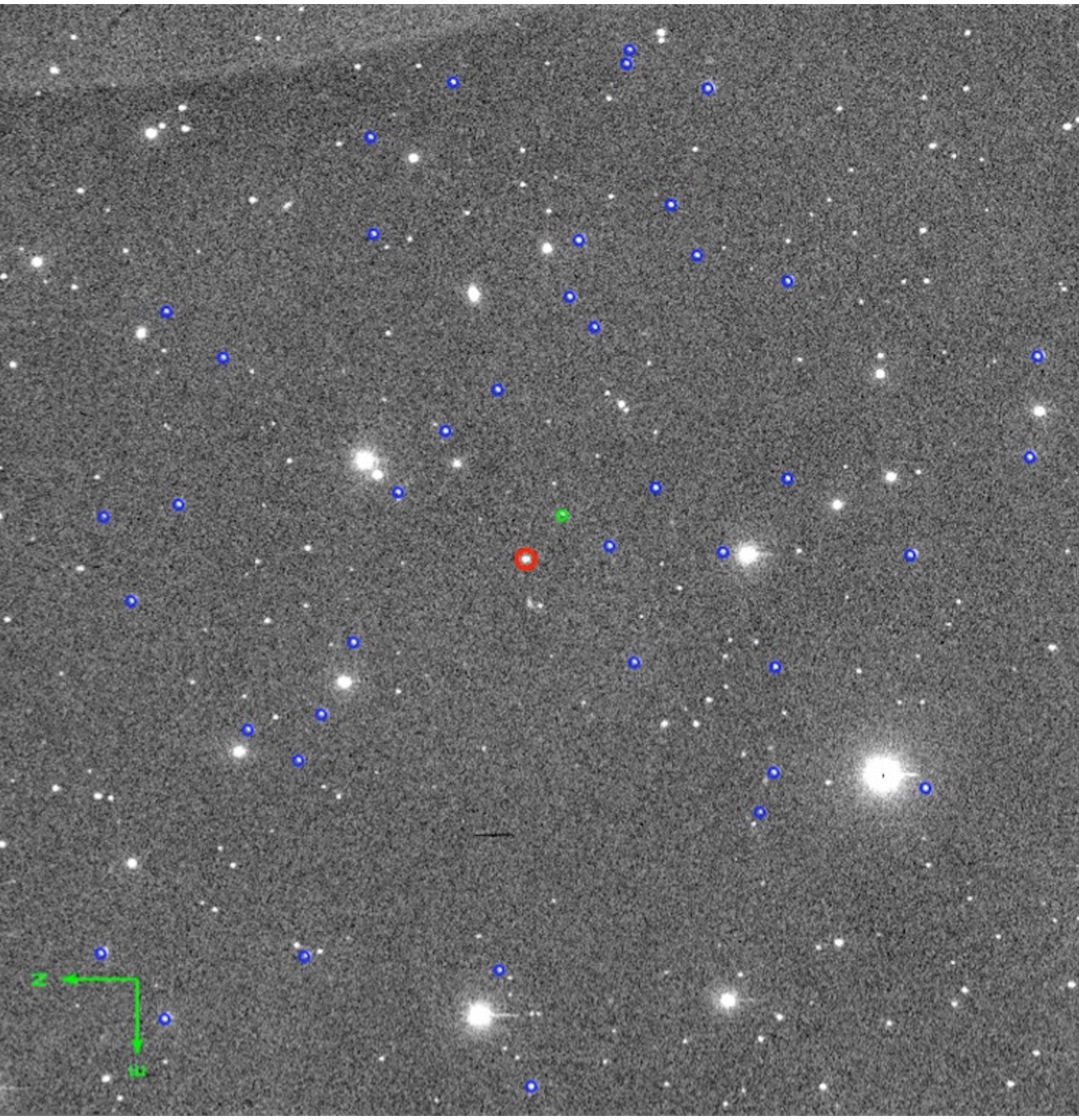}
	\caption{B-band image of PG 0934+013 (green circle) obtained at the LOAO 1-m telescope. Comparison stars, which were used in the differential photometry, are denoted with blue circles while the comparison star in the slit for the SALT spectroscopy is marked with a red circle. The FoV is 28'$\times$28'. 
	}
\end{figure}

\subsection{Photometry}

We carried out imaging observations for the broad band photometry, using the \textit{B}-band filter. In total we obtained \textit{B}-band photometry from 80 epochs during the campaign. We started the photometry campaign on December 6th 2016 with a $\sim$4 day cadence, 
using the Faulkes Telescope North (FTN) 2-m in Hawaii, which is one of the Las Cumbres Observatory Global Telescopes (LCOGT).  
The campaign lasted until April 11th 2013 when the the target became unobservable. We will call this campaign period as Year 1.
Additionally, we obtained 3 epochs in October 2012.
In the second year we continued the campaign from December 10th 2013 to May 12th 2014 with a $\sim$5 day cadence (Year 2), using the Mt. Lemon Optical Astronomy Observatory (LOAO) 1-m telescope in  Arizona. During the second year campaign, we also used the Faulkes Telescope South (FTS) for seven epochs between December 2013 and January 2014 to assist the LOAO campaign. However, we mainly used the LOAO data for the reverberation analysis for Year 2. 

Both LCOGT 2-m telescopes (FTN and FTS) have the same detector: 4K$\times$4K Fairchild CCD with a 15$\mu$m pixel size, providing a $10\arcmin\times10\arcmin$ field of view (FoV). The images were taken in 2$\times$2 pixel binning mode with a spatial scale of 0$\farcs$304 pixel$^{-1}$.
LOAO 1-m telescope has a 4K$\times$4K E2V CCD detector with a 15$\mu$m pixel size, providing a $28\arcmin\times28\arcmin$ FoV. 
We used 2$\times$2 pixel binning mode, which corresponds to a pixel scale of 0$\farcs$796 pixel$^{-1}$. 

We determined the exposure time for \textit{B}-band photometry to obtain high quality data. 
Typical total exposure time was 180 and 480 seconds, respectively for FT and LOAO, which is divided into 3 to 4 exposures. 
The quality of images is good enough for differential photometry except for a couple of epochs when the sky condition was bad (e.g., seeing, cirrus, etc). 

FT data were delivered to us after pre-process (i.e., bias subtraction and flat fielding), while for the LOAO data, we preprocessed the data following the standard IRAF procedure.
We combined individual exposures using median combine for each night, generating one image per epoch. 
Cosmic-rays were removed with the LA-Cosmic task \citep{vanDokkum2001}. 
Photometric measurements for the target AGN and comparison stars in the field were carried out using SExtractor (Bertin \& Arnouts 1996).
We performed aperture photometry using various aperture sizes ranging from 1$\arcsec$ to 20$\arcsec$ diameters, 
to test the seeing effect and the reliability of the flux measurements. 
We found that the aperture diameter of 7$\arcsec$ provides an optimal aperture with a minimum systematic variation. 
Thus, we used a 7$\arcsec$ aperture for all sources in the FoV.

Since our goal is to perform differential photometry for measuring the relative flux change, we used a number of stars in the FoV to 
calculate zero-points of each epoch assuming these stars are not variable. For this step, we selected bright stars, which have the uncertainty of the instrumental magnitude less than 1\%. All of these stars turn out to have B $<$ 17 mag.
Typically $\sim$10 and $\sim$40 stars were used as comparison stars, respectively for LCOGT and LOAO data (see Figure 1).
 
We chose the best epoch when the sky condition was close to photometric and calculated the magnitude difference ($\Delta$mag.) of each star between the standard epoch and each epoch. The average of $\Delta$mag. provides the zero point of given epochs while the standard deviation of $\Delta$mag. represents the systematic uncertainty of the zero points, which is typically 0.03 mag for LCOGT and 0.02 mag for LOAO photometry.
Once we obtained the zero point for each epoch using the $\Delta$mag. method, we rescaled all magnitude measurements using the zero point. 
As a consistency check, we calculated the rms variability of each comparison star, and found that flux variability is less than 
0.02-0.03 mag during the campaign, suggesting that the $\Delta$mag. method provided reliable calibration within a few percent systematic uncertainty. At the same time, we confirmed that these comparison stars were not variable based on the results from the differential photometry.

As a standard star, we used a bright star ID 0910-0183676 (R.A.=09:36:59.60, Dec.=+01:03:40.5), which is close to the target AGN, 
and adopted the \textit{B}-band magnitude B1=15.72 from USNO B1.0 catalog. Assuming that \textit{B} magnitude of LCOGT and LOAO are similar to B1 magnitude of USNO B1.0 catalog, we then calibrated all magnitude measurements. Since our aim is to measure the time lag between photometry and spectroscopy, this simple 
calibration is acceptable without affecting the time-lag measurements.  
Table 1 lists the calibrated \textit{B}-band magnitude of PG 0934+013. The quoted uncertainty was derived by combining the measurement
error (i.e., standard photometric error) from SExtractor and the systematic error based on the rms scatter of 
the zero point of individual comparison stars.
 
\subsection{Spectroscopy}

Spectroscopic observations were performed during 46 nights scheduled between December 20th 2012 and April 29th 2014, along with a seven month gap between Year 1 and Year 2 due to the lack of observability. Spectroscopic observations were carried out  with the Robert Stobie Spectrograph \citep[e.g.,][]{Burgh2003,Kobulnicky2003} at the SALT \citep{Buckley2006}, which was entirely performed in service mode. 
The \textit{B}-band magnitude of the target was reported as 16.3 by \cite{Schmidt&Green83} and we calculated the \Hb\ lag
based on the monochromatic luminosity at 5100\AA\ measured from the observed flux in the SDSS spectrum, using the size-luminosity relation from \cite{Bentz2013}. The expected lag is $\sim$25 days although its value has large uncertainty due to the variability and the contribution from the host galaxy emission to the observed continuum flux. Assuming that the actual lag would be close to 25 days, we chose a $\sim$five day cadence for the monitoring program, to obtain sufficient time resolution for cross correlation analysis.

We used a spectral setup with the PG2300 grating centered at 5136 \AA, covering a spectral range between 4636 \AA\ and 5636 \AA, in which the \Hb\ and \OIII\ emission lines are located between the two gaps in the CCD. To minimize the slit loss we used slit width 4\arcsec.
The spectral resolution is $R=1040$ at the center of the spectral range while the spatial scale is 0\farcs254 pixel$^{-1}$ in the 2$\times$2 binning mode. We used the position angle from North to East 50.5$^{\circ}$ to include a comparison star in the slit for experimenting flux calibration. However, we did not attempt to perform flux calibration based on the comparison star in the slit since the SALT flux 
calibration has large uncertainty due to the vignetting, which changes at each epoch.
During the first 5 epochs in Year 1, the spectral set-up was not optimal, hence we adjusted the grating angle to move a part of blue wing of \Hb\ emission line out of the CCD gap in the spectral range. Also, we increased the slit width from 2\arcsec\ to 4\arcsec\ for the rest of the campaign. Note that these 5 epochs were not included in the analysis.
 
We determined the exposure time using the RSS simulator\footnote{http://astronomers.salt.ac.za/software/rss-simulator/} to achieve high S/N$>$20. We initially used a total of 360 second exposure per epoch, then increased the exposure time to 540 second, which was divided into 2 readouts
(see Table 3). Arc lamp and dome flat images were taken each night for standard calibration. 
Spectroscopic data were processed by the SALT PyRAF package \citep[PySALT;][]{Crawford2010}, which includes bias subtraction, flat-fielding, gain correction, cross-talk correction, and amplifier mosaicking. Cosmic rays were cleaned using the LA-Cosmic routine \citep{vanDokkum2001}. Wavelength calibration was performed by identifying the arc lines from arc spectra and rectifying spectral image based on a polynomial fit. 
The spectral images were combined using a median combine. 
Then, we extracted one dimensional spectra with a 7.62\arcsec\ aperture size, 
using the procedures in the PySALT package. The median S/N per pixel at 5100 \AA\ is $\sim$37 (see Table~\ref{table2}).

Flux calibration was initially performed using spectrophotometric standard stars, which were observed during 11 nights out of the total 46 nights. Note that standard stars are typically observed on a weekly basis at the SALT to constrain the response function over the
spectral range. For the nights when standard stars were not observed, we adopted the flux calibration from adjacent observing dates
as we assumed the response function would be similar for a given small spectral range (i.e., including \Hb\ and \OIII). 
Since the uncertainty of the flux calibration of the SALT spectroscopy is relatively large, we recalibrated each spectrum based on the narrow emission line [OIII] \lam5007, which is expected to be non-variable in the time scale of our observing campaign. 
For this process, we followed the procedure outlined by \citet{vanGroningenWanders1992}, by shifting and rescaling individual spectra with respect to the reference spectrum, which was averaged from all spectra obtained during the campaign. We first decomposed each spectrum as described in the next Section and measured the flux of the \OIII\ line using the best-fit model. Then, we rescaled each spectrum to match the \OIII\ flux. 

In addition, we also used the flux calibration code presented by \citet{Fausnaugh2016}, who used the same scheme of \citet{vanGroningenWanders1992}. The code by \citet{Fausnaugh2016} is less dependent on the spectral resolution, which may vary night-to-night due to the varying seeing, and provides a better smoothing kernel with Gauss-Hermite polynomials for matching the spectral resolution. The code uses a Bayesian formalism for fitting the line profiles and estimating parameter uncertainties. As a consistency check of the flux recalibration, we employed this code to rescale individual spectra and obtained the \Hb\ light curve and the \Hb\ lag, which were consistent with the measurements based on our own rescaling method (see more details in Section 3.2).

\section{Analysis}

\subsection{Multi-component Fitting}
\begin{figure}
	\includegraphics[width = 0.49\textwidth, height=0.3\textwidth]{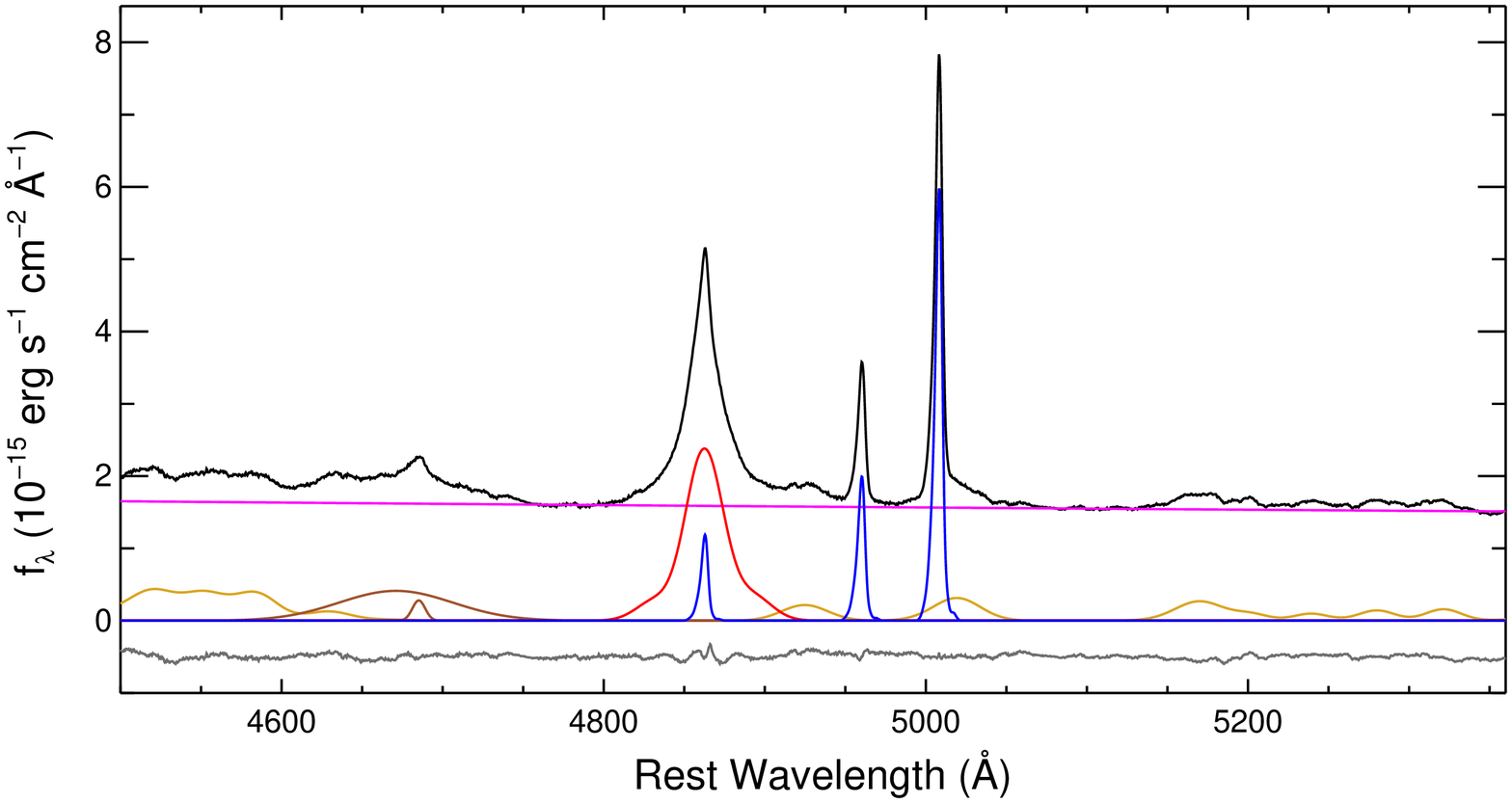}
	\caption{Example of the spectral decomposition with multi-components. The mean spectrum (black) is modeled with a power-law continuum (magenta), a broad component of \Hb\ (red), a narrow component of \Hb\ (blue), \OIII\ \lam\lam 4959, 5007 (blue), broad and narrow components of \HeII\ (brown), 
	and \FeII\ emission blends (yellow).  At the bottom, the residual of the fit is presented with a grey line.
		\label{fig1}}
\end{figure}

We performed multi-component spectral fitting to measure the fluxes and widths of the major emission lines, e.g., \Hb, \OIII, and \HeII.
Each spectrum was decomposed into AGN featureless continuum, the \FeII\ emission blends, \Hb,  \OIII, and \HeII\ emission lines \citep[see][]{Woo2006, Park2012}. The Levenberg-Marquardt routines in the IDL \texttt{mpfit} package \citep{Markwardt2009} were adopted to determine the best model based on $\chi^2$ minimization. We used a power-law component to represent featureless AGN continuum while the stellar component becomes negligible in the fitting process in most cases except for several epochs with high S/N (i.e., $>$ $\sim$50) when stellar absorption lines were noticeable. Generally, stellar lines were very weak although we see the sign of stellar absorption lines (i.e., Mgb triplet and Fe 5270 \AA) in the residual (see Figure~\ref{fig1}).

\FeII\ emission blends were initially modeled with the \FeII\ template from Boroson \& Green (1992). However, due to the strong \FeII\ emission features between \Hb\ and \OIII\ \lam4696, and redward of \OIII\ \lam5007, this template failed to provide acceptable fitting results. Instead, we used the \FeII\ template provided by \cite{Kovacevic2010}, who divided the \FeII\ complex into five groups, so that various flux ratios among these five groups can be applied. We find that this template provides better fit although the strong \FeII\ emission features between \Hb\ and \OIII\ \lam4696 cannot be subtracted entirely. Since we mainly focus on the \Hb\ line flux measurements, we accept these results for our variability analysis. 
In the case of the \HeII\  line, we modeled the line profile with broad and narrow components using two Gaussian components, which provided successful results. For the \Hb\ \lam4861 and \OIII\ \lam\lam4959, 5007 lines, we used a more complex fit. A tenth-order Gauss-Hermite series \citep{Marel1994} was employed to fit \OIII\ \lam5007 in the range of 4979-5022 \AA. The \OIII\ \lam4959 line was modeled with the same velocity profile and the flux scale of 1/3. The \Hb\ narrow component was also constructed using the \OIII\ profile 
after scaling to match the \Hb\ narrow component. A typical flux scale between narrow \Hb\ and \OIII\ \lam5007 is 0.19.
Finally, the \Hb\ broad component was modeled using a tenth-order Gauss-Hermite functions in the range of 4791-4931 \AA. 
We fit all aforementioned components simultaneously and measured the line fluxes based on the best model \citep[cf.][]{Woo2006,Park2012,Barth2015}. 
In Figure~\ref{fig1}, we present an example of the best-fit model.
Note that since it is difficult to separate the narrow and broad components in the \Hb\ line profile as the flux ratio between \Hb\ and \OIII\ changes epoch to epoch in the fitting process, we decided to use the total flux of the \Hb\ line profile for time lag analysis, in order to avoid additional systematic uncertainties. Thus, the variability of \Hb\ is slightly underestimated since the constant narrow \Hb\ flux is added. 

\begin{figure*}
	\centering
	\includegraphics[width=0.9\textwidth]{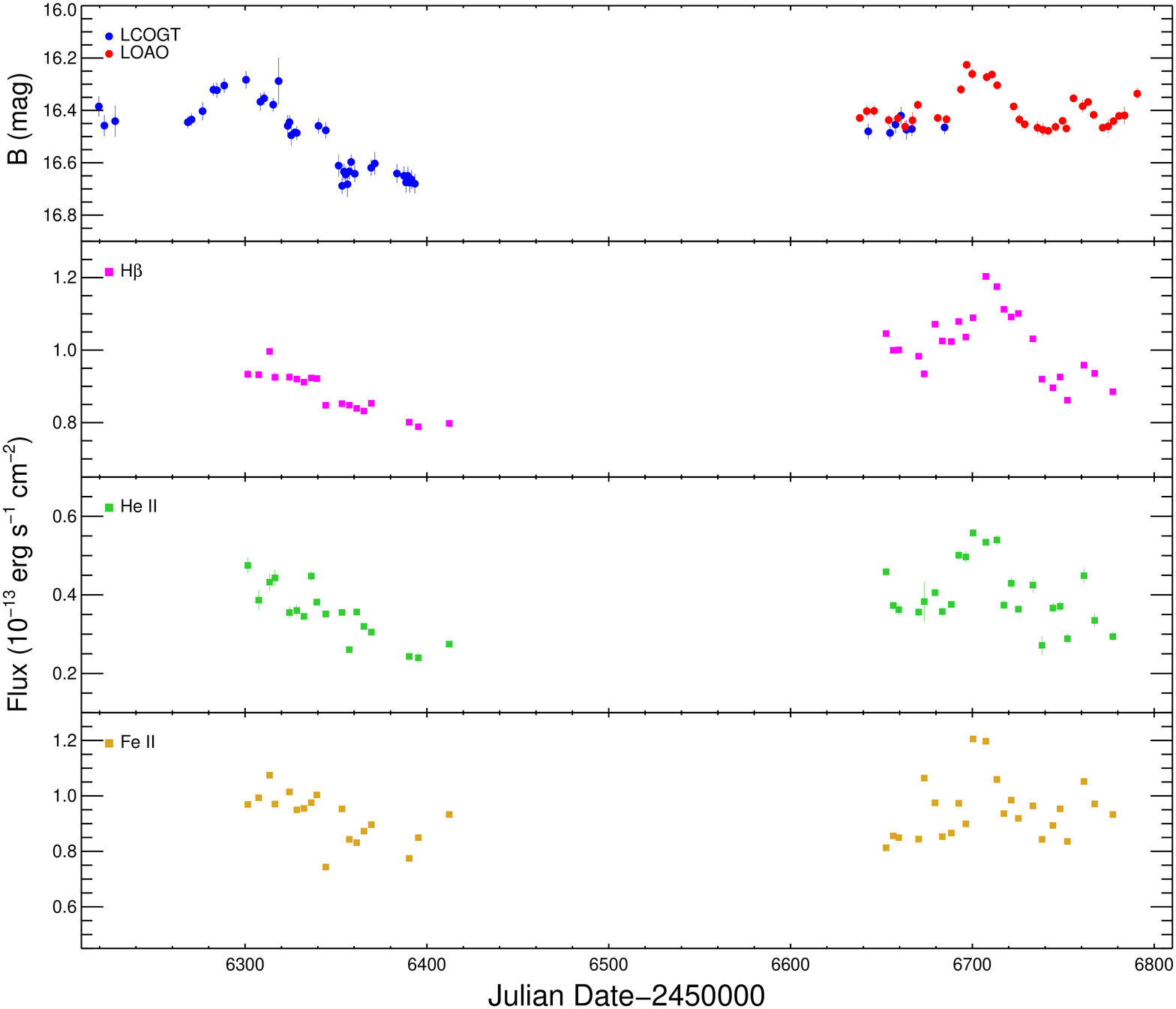}
	\caption{Light curves of the \textit{B} band, the \Hb , \HeII , and \FeII\ emission lines (from top to bottom).
	}
\end{figure*}

\begin{table}
	\centering
	\caption{\textit{B}-band Magnitude}
	\begin{tabular}{cccc}
		\tableline\tableline
		\multicolumn{2}{c}{Year 1 (2012-2013)} & \multicolumn{2}{c}{Year 2 (2013-2014)} \\
		\cline{1-2} \cline{3-4}
		JD &  \textit{B} & JD &  \textit{B} \\
		(days) & (mag) & (days) & (mag) \\
		(1) & (2) & (3) & (4) \\
		\tableline
   6219.586$\rm^a$  &   16.39  $\pm$    0.04 &    6638.050  &   16.43  $\pm$    0.01  			\\
   6222.612$\rm^a$  &   16.46  $\pm$    0.04 &    6641.947  &   16.40  $\pm$    0.02          \\
   6228.556$\rm^a$  &   16.44  $\pm$    0.06 &    6642.717$\rm^b$  &   16.48  $\pm$    0.03   \\
   6268.528  &   16.45  $\pm$    0.02        &    6645.930  &   16.40  $\pm$    0.02          \\
   6270.530  &   16.43  $\pm$    0.02        &    6654.017  &   16.44  $\pm$    0.02          \\
   6276.618  &   16.40  $\pm$    0.03        &    6654.719$\rm^b$  &   16.49  $\pm$    0.03   \\
   6282.601  &   16.32  $\pm$    0.02        &    6657.722$\rm^b$  &   16.45  $\pm$    0.03   \\
   6284.527  &   16.32  $\pm$    0.03        &    6659.052  &   16.43  $\pm$    0.02          \\
   6288.545  &   16.30  $\pm$    0.03        &    6660.753$\rm^b$  &   16.42  $\pm$    0.03   \\
   6300.529  &   16.28  $\pm$    0.03        &    6663.034  &   16.46  $\pm$    0.02          \\
   6308.450  &   16.37  $\pm$    0.03        &    6663.750$\rm^b$  &   16.47  $\pm$    0.04   \\
   6310.464  &   16.35  $\pm$    0.03        &    6666.734$\rm^b$  &   16.47  $\pm$    0.03   \\
   6315.473  &   16.38  $\pm$    0.03        &    6667.051  &   16.44  $\pm$    0.03          \\
   6318.445  &   16.29  $\pm$    0.09        &    6669.991  &   16.38  $\pm$    0.02          \\
   6323.469  &   16.46  $\pm$    0.04        &    6680.954  &   16.43  $\pm$    0.02          \\
   6324.400  &   16.45  $\pm$    0.03        &    6684.711$\rm^b$  &   16.46  $\pm$    0.03   \\
   6325.400  &   16.50  $\pm$    0.04        &    6685.719  &   16.43  $\pm$    0.02          \\
   6327.379  &   16.48  $\pm$    0.01        &    6693.718  &   16.32  $\pm$    0.02          \\
   6328.401  &   16.49  $\pm$    0.03        &    6696.858  &   16.23  $\pm$    0.02          \\
   6340.386  &   16.46  $\pm$    0.03        &    6699.896  &   16.26  $\pm$    0.02          \\
   6344.383  &   16.48  $\pm$    0.03        &    6707.892  &   16.27  $\pm$    0.02          \\
   6351.457  &   16.61  $\pm$    0.04        &    6710.767  &   16.26  $\pm$    0.01          \\
   6353.379  &   16.69  $\pm$    0.03        &    6713.646  &   16.30  $\pm$    0.02          \\
   6354.383  &   16.63  $\pm$    0.03        &    6722.716  &   16.39  $\pm$    0.01          \\
   6355.296  &   16.64  $\pm$    0.03        &    6725.855  &   16.43  $\pm$    0.02          \\
   6356.315  &   16.68  $\pm$    0.05        &    6728.776  &   16.45  $\pm$    0.02          \\
   6357.335  &   16.63  $\pm$    0.03        &    6735.794  &   16.47  $\pm$    0.02		  \\
   6358.336  &   16.60  $\pm$    0.03        &    6738.806  &   16.47  $\pm$    0.02          \\
   6360.320  &   16.64  $\pm$    0.03        &    6741.732  &   16.48  $\pm$    0.01          \\
   6369.366  &   16.62  $\pm$    0.03        &    6745.740  &   16.46  $\pm$    0.02          \\
   6371.333  &   16.60  $\pm$    0.04        &    6749.612  &   16.44  $\pm$    0.01          \\
   6383.522  &   16.64  $\pm$    0.04        &    6751.653  &   16.47  $\pm$    0.01          \\
   6387.403  &   16.65  $\pm$    0.04        &    6755.625  &   16.35  $\pm$    0.01          \\
   6388.536  &   16.68  $\pm$    0.04        &    6760.658  &   16.38  $\pm$    0.02          \\
   6389.536  &   16.65  $\pm$    0.03        &    6763.648  &   16.37  $\pm$    0.01          \\
   6390.528  &   16.68  $\pm$    0.04        &    6766.699  &   16.42  $\pm$    0.01          \\
   6391.572  &   16.66  $\pm$    0.04        &    6771.667  &   16.47  $\pm$    0.01          \\
   6393.377  &   16.68  $\pm$    0.04        &    6774.686  &   16.46  $\pm$    0.03          \\
   &                               &    6777.653  &   16.44  $\pm$    0.01          \\
   &                               &    6780.670  &   16.42  $\pm$    0.01          \\
   &                               &    6783.659  &   16.42  $\pm$    0.03          \\
   &                               &    6790.655  &   16.34  $\pm$    0.02 			\\
	\tableline
	\end{tabular}
	\label{table1}
		\tablecomments{
			Col. 1: Julian date (-2,450,000).
			Col. 2: \textit{B}-band magnitude in Year 1.
			Col. 3: Julian date (-2,450,000).
			Col. 4: \textit{B}-band magnitude in Year 2. \\
			{$\rm^a$ These epochs were not used for the CCF analysis because of a large time gap.} \\
			{$\rm^b$ The data of these epochs were obtained at the FTS in Year 2, and excluded for the CCF analysis.}
		}

\end{table}

\begin{table}
	\centering
	\caption{Emission line fluxes}
	\begin{tabular}{cccccc}
		\tableline\tableline
		JD & $t_{\rm exp}$ & S/N & $f_{\rm H\beta}$  &  $f_{\rm He {\small II}}$ & $f_{\rm Fe {\small II}}$  \\
		(days) & (s) &  & \multicolumn{3}{c}{ ($\rm 10^{-15}~erg~s^{-1}~cm^{-2}$) } \\
		(1) & (2) & (3) & (4) & (5) & (6) \\
		\tableline
  6301.511  &  360  &  28 &  93.36  $\pm$    1.27  &   47.49  $\pm$    2.10  &   96.92  $\pm$    0.27   \\
  6307.492  &  360  &  41 &  93.23  $\pm$    1.12  &   38.67  $\pm$    2.59  &   99.34  $\pm$    0.29   \\
  6313.475  &  360  &  45 &  99.66  $\pm$    0.96  &   43.24  $\pm$    2.16  &  107.46  $\pm$    0.28   \\
  6316.448  &  360  &  29 &  92.52  $\pm$    1.37  &   44.32  $\pm$    2.06  &   97.05  $\pm$    0.34   \\
  6324.417  &  540  &  30 &  92.54  $\pm$    1.14  &   35.51  $\pm$    1.47  &  101.44  $\pm$    0.19   \\
  6328.446  &  540  &  52 &  92.01  $\pm$    0.98  &   36.01  $\pm$    1.47  &   94.97  $\pm$    0.21   \\
  6332.388  &  540  &  55 &  91.17  $\pm$    0.53  &   34.52  $\pm$    0.74  &   95.47  $\pm$    0.13   \\
  6336.392  &  540  &  62 &  92.34  $\pm$    0.53  &   44.79  $\pm$    1.31  &   97.55  $\pm$    0.13   \\
  6339.444  &  540  &  66 &  92.15  $\pm$    0.49  &   38.16  $\pm$    0.86  &  100.36  $\pm$    0.12   \\
  6344.352  &  540  &  23 &  84.79  $\pm$    0.36  &   35.14  $\pm$    0.56  &   74.33  $\pm$    0.09   \\
  6353.339  &  540  &  29 &  85.23  $\pm$    0.44  &   35.53  $\pm$    0.83  &   95.32  $\pm$    0.11   \\
  6357.323  &  540  &  43 &  84.82  $\pm$    0.35  &   26.04  $\pm$    0.65  &   84.32  $\pm$    0.11   \\
  6361.360  &  540  &  41 &  83.92  $\pm$    0.47  &   35.64  $\pm$    0.87  &   83.15  $\pm$    0.11   \\
  6365.350  &  540  &  53 &  83.19  $\pm$    0.39  &   31.97  $\pm$    0.84  &   87.30  $\pm$    0.12   \\
  6379.380  &  540  &  18 &  85.33  $\pm$    0.44  &   30.51  $\pm$    0.99  &   89.63  $\pm$    0.14   \\
  6390.296  &  540  &  46 &  80.13  $\pm$    0.35  &   24.32  $\pm$    0.76  &   77.48  $\pm$    0.10   \\
  6395.304  &  540  &  32 &  78.86  $\pm$    0.41  &   23.98  $\pm$    1.26  &   84.95  $\pm$    0.12   \\
  6412.254  &  540  &  51 &  79.80  $\pm$    0.40  &   27.47  $\pm$    0.87  &   93.28  $\pm$    0.12   \\
  6652.544  &  540  &  28 & 104.56  $\pm$    0.66  &   45.87  $\pm$    1.12  &   81.29  $\pm$    0.15   \\
  6656.532  &  540  &  49 &  99.97  $\pm$    0.61  &   37.30  $\pm$    1.14  &   85.58  $\pm$    0.19   \\
  6659.514  &  540  &  46 & 100.06  $\pm$    0.76  &   36.22  $\pm$    1.34  &   84.96  $\pm$    0.15   \\
  6670.473  &  720  &  27 &  98.31  $\pm$    0.89  &   35.62  $\pm$    1.06  &   84.40  $\pm$    0.16   \\
  6673.509  &  540  &  20 &  93.45  $\pm$    1.04  &   38.29  $\pm$    5.12  &  106.41  $\pm$    0.15   \\
  6679.462  & 1080  &  41 & 107.15  $\pm$    0.47  &   40.56  $\pm$    0.93  &   97.52  $\pm$    0.11   \\
  6683.471  &  540  &  39 & 102.51  $\pm$    0.57  &   35.75  $\pm$    1.09  &   85.30  $\pm$    0.16   \\
  6688.420  &  540  &  33 & 102.37  $\pm$    0.59  &   37.56  $\pm$    0.91  &   86.60  $\pm$    0.15   \\
  6692.489  &  540  &  39 & 107.88  $\pm$    0.60  &   50.11  $\pm$    1.12  &   97.35  $\pm$    0.13   \\
  6696.406  &  540  &  44 & 103.60  $\pm$    0.75  &   49.66  $\pm$    1.46  &   89.93  $\pm$    0.14   \\
  6700.379  &  540  &  23 & 108.92  $\pm$    0.64  &   55.76  $\pm$    1.06  &  120.54  $\pm$    0.17   \\
  6707.408  &  540  &  22 & 120.33  $\pm$    0.45  &   53.41  $\pm$    0.85  &  119.73  $\pm$    0.11   \\
  6713.476  &  540  &  44 & 117.51  $\pm$    0.74  &   53.98  $\pm$    1.31  &  105.95  $\pm$    0.18   \\
  6717.349  &  540  &  40 & 111.25  $\pm$    0.65  &   37.39  $\pm$    1.20  &   93.64  $\pm$    0.17   \\
  6721.338  &  540  &  43 & 109.14  $\pm$    0.77  &   42.95  $\pm$    1.34  &   98.48  $\pm$    0.16   \\
  6725.318  &  540  &  23 & 110.08  $\pm$    0.61  &   36.35  $\pm$    0.93  &   91.90  $\pm$    0.17   \\
  6733.303  &  540  &  13 & 103.12  $\pm$    0.54  &   42.48  $\pm$    1.93  &   96.41  $\pm$    0.16   \\
  6738.301  &  540  &  43 &  91.99  $\pm$    0.83  &   27.16  $\pm$    2.36  &   84.31  $\pm$    0.23   \\
  6744.297  &  540  &  43 &  89.61  $\pm$    0.76  &   36.65  $\pm$    1.39  &   89.33  $\pm$    0.18   \\
  6748.266  &  540  &  31 &  92.60  $\pm$    0.62  &   37.10  $\pm$    1.37  &   95.36  $\pm$    0.17   \\
  6752.270  &  540  &  37 &  86.17  $\pm$    0.74  &   28.85  $\pm$    1.32  &   83.60  $\pm$    0.25   \\
  6761.359  &  540  &  15 &  95.85  $\pm$    0.83  &   44.90  $\pm$    1.90  &  105.22  $\pm$    0.13   \\
  6767.261  &  540  &  46 &  93.58  $\pm$    1.05  &   33.51  $\pm$    1.72  &   97.11  $\pm$    0.18   \\
  6777.244  &  540  &  33 &  88.52  $\pm$    0.66  &   29.40  $\pm$    1.17  &   93.31  $\pm$    0.16   \\
  
		\tableline
	\end{tabular}
	\label{table2}
	\tablecomments{
		Col. 1: Julian date (-2,450,000).
		Col. 2: Exposure time.
		Col. 3: Signal-to-noise per pixel averaged in the range of 5090-5110 \AA.
		Col. 4: Flux of \Hb\ emission line.
		Col. 5: Flux of \HeII\ emission line.
		Col. 6: Flux of \FeII\ emission blends.
		}

\end{table}

\begin{table}
	\centering
	\caption{Variability of continuum and emission lines}
	\begin{tabular}{lcc}
		\tableline\tableline
		    & $F_{\rm var}$  & $R_{\rm max}$ \\
		\tableline
		\textit{B}-band  & 0.11 & 1.53 $\pm$ 0.05 \\
		\Hb\  &  0.11 &  1.53 $\pm$ 0.01 \\
		\HeII\  &  0.22 & 2.33 $\pm$ 0.13 \\
		\FeII\ & 0.10 & 1.62 $\pm$ 0.01 \\
		\tableline
	\end{tabular}
	\label{table3}
\end{table}

\subsection{Time Lag Measurements}

In Figure~3 we present the light curves of \textit{B}-band photometry, \Hb , \HeII\ emission lines, and \FeII\ emission blends. The \textit{B}-band magnitude measurements are listed in Table~\ref{table1} while the spectroscopic flux measurements of each emission line are listed in~Table \ref{table2}. 
We quantified the variability of each light curve using $F_{\rm var}$,
which is defined as
\begin{equation}\label{eq:var}
F_{\rm var}=\frac{\sqrt{\sigma^2 - \langle \delta^2 \rangle}}{\langle f \rangle}\ 
\end{equation}
where $\sigma^2$ is the variance of the fluxes, $\langle \delta^2 \rangle$ is the rms uncertainty, and $\langle f \rangle$ is the mean of the observed fluxes \citep{RodriguezPascul1997,Edelson2002}. 
We find that $F_{\rm var}$ is larger than 0.1 for all light curves, indicating high variability as listed in Table~\ref{table3}. The \HeII\ emission line shows the highest variability amplitude, which is roughly two times higher than those of \Hb\ and \FeII.
We also calculated $R_{\rm max}$, which is the ratio of the maximum and minimum fluxes in the light curve, 
finding that the fluxes of the continuum and \Hb\ and \FeII\ have increased by 50-60\% during the campaign, while
\HeII\ shows more than a factor of 2 increase. 

Using the light curves presented in Figure 3, we calculated the cross correlation function (CCF) to measure the time lag between continuum and each emission line, following the procedure described by \citet{Peterson2004}. Due to the seven month gap between two monitoring campaigns, we separately determined a CCF for each year. Since the continuum and emission line data were unevenly sampled, we first re-distributed the data by interpolating \textit{B}-band light curve with respect to the emission line sampling, then derived the correlation coefficient.
In the same way, we interpolated the light curve of each emission line with respect to continuum sampling, then calculated the CCF \citep{GaskellPeterson1987,WhitePeterson1994,Peterson2004}. The CCF was computed for time lags ranging from $-$20 to 60 days with increments of 0.25 days. Once we obtained two CCF results, we used the averaged CCF for determining the time lags and their uncertainties.
We determined time lags using two different definitions: \lagpeak\ using the peak of CCF, and \lagcen\ using the centroid of CCF, which was calculated using the points with $\rm CCF \geq 0.8 \times CCF_{\rm peak}$. 

The final time lags and their uncertainties were determined by using Monte Carlo realization ($N$=10,000) based on flux randomization and random subset sampling (FR/RSS) method \citep{GaskellPeterson1987,WhitePeterson1994,Peterson2004}. We randomly selected the same number of data points from the light curve, allowing duplication. If one epoch was selected $m$ times, then we decreased the uncertainty 
of the flux by $m^{1/2}$. Using the selected epoch, then we simulated mock light curves by adding Gaussian random noise based on the flux errors at each epoch, and computed the CCF of the generated continuum and emission light curves. 
From the distribution of the lag measurements, we adopted the mean value as the final time lag, and the difference between the mean and the highest (lowest) 1$\sigma$ value in the distribution of CCF was adopted for uncertainty.

\begin{figure}
	\includegraphics[width = 0.43\textwidth]{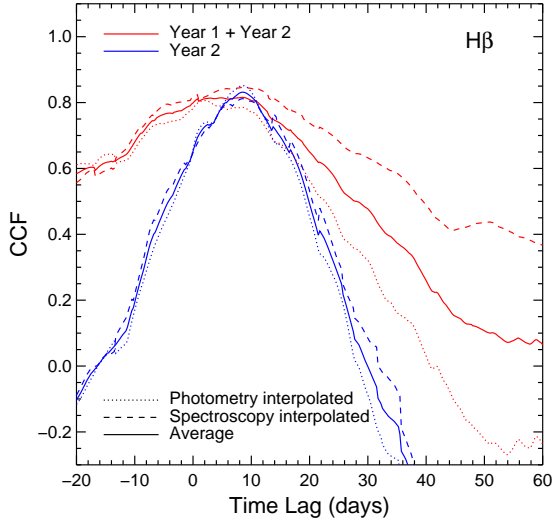}
	\caption{Cross-correlation function between the \textit{B}-band continuum and \Hb\ emission line, respectively for Year 1 + Year 2 (red) and year 2 (blue). CCF values are calculated by interpolating photometry data (dotted) or spectroscopy data (dashed) data, respectively. The average 
	of the two CCF results are denoted with solid lines.
		\label{fig3}}
\end{figure}

Figure~\ref{fig3} presents the results of CCF calculation between the continuum and the \Hb\ emission line. We measured the time lag  \lagpeak\ = $8.52^{+2.23}_{-2.27}$ days, and \lagcen\ = $8.46^{+2.08}_{-2.14}$ days in the observed frame from Year 2 data while we were not able to determine the time lag based on Year 1 data due to the monotonic decrease of the \Hb\ flux. 
For a consistency check for the interpolation process and cross correlation analysis, we used the JAVELIN code (Zu et al. 2013) to perform CCF calculation based on Year 2 data, obtaining \Hb\ time lag $\tau_{\rm JAVELIN}$ = $8.32 \pm 2.04 $ days in the observed frame, which is consistent within the uncertainty. 

Considering the uncertainties due to the flux recalibration based on \OIII, we also tested how the lag measurement changes if we use the rescaled spectra using the code by \citet{Fausnaugh2016}. We found that when we decomposed each spectrum and used the emission line spectrum for flux rescaling after subtracting FeII complex and the power-law component, we obtained a very similar \Hb\ light curve except for a few epochs, which showed the lowest S/N among all epochs. Based on the Monte Carlo simulations, we obtained the \Hb\ time lag \lagcen\ = $8.84^{+2.26}_{-2.35}$ days, which is consistent with the results based on our own flux rescaling
method within the uncertainties. Thus, we confirm that the difference of the flux rescaling method is insignificant for time lag measurements. In contrast, when we used the original spectra without subtracting FeII complex and the power-law component, we obtained a very different light curve and failed to obtain a reliable lag measurement. We suspect that these results are due to the fact that the \FeII\ feature, that is blended with \OIII\ $\lambda$5007, is very strong and its variability may cause systematic uncertainties in measuring \OIII\ flux in each single-epoch spectrum if the decomposition is not properly performed. 

\begin{figure}
	\includegraphics[width = 0.43\textwidth]{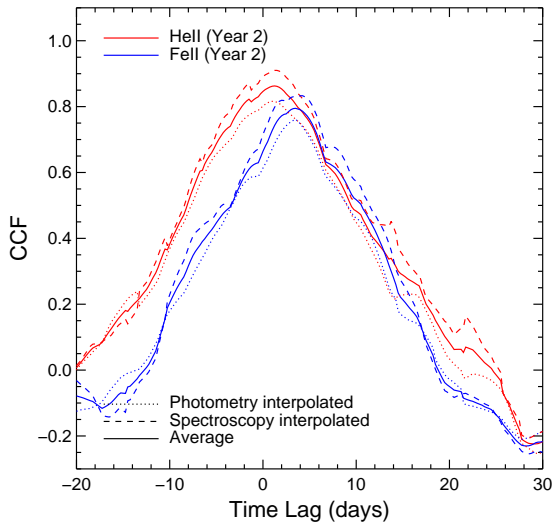}
	\caption{Cross-correlation function between the \textit{B}-band continuum and \HeII\ emission line (red) and \FeII\ emission blends (blue) 
	based on Year 2 data.
		\label{fig4}}
\end{figure}

We also investigated the time lag of the \HeII\ line and \FeII\ blends (see Table 4).
While the distribution of the CCF of \HeII\ was not well-constrained with Year 1 data, we were able to measure the time lags of \HeII\  
for Year 2 (Figure~\ref{fig4}). These results are expected in Figure~3 as the light curve of \HeII\ in Year 1 shows monotonic decrease without a strong bump. 
We determined the time lag of \HeII\ in the observed frame as \lagpeak\ = $0.92^{+1.58}_{-2.42}$ days and \lagcen\ = $0.84^{+2.05}_{-2.19}$ days based on Year 2 data.
Although the measured \HeII\ lag is consistent with zero within the uncertainty, it suggests that the \HeII\ lag is much shorter than the \Hb\ lag. 
This result is consistent with the fact that \HeII\ is strongly variable  and promptly responding to continuum emission as also noted by previous reverberation mapping results \citep{PetersonFerland1986,Kollatschny2003,Pei2014,Barth2015}.

For \FeII, the light curve in Year 1 appears to have a similar trend to \Hb, but the CCF was not well-defined.
Instead, we determined the \FeII\ lag \lagpeak\ = $3.51^{+1.49}_{-2.76}$ days and \lagcen\ = $3.74^{+1.97}_{-2.62}$ days in 
the observed frame based on Year 2 data. This result  suggests that 
\FeII\ emission appears to be located in an inner region than \Hb\ \citep[cf.][]{Barth2013,Hu2015}. However, due to the large
uncertainty of the lag and the relatively weak CCF result (see Fig. 4), further analysis based on a better quality light curve is needed
to derive more reliable conclusion.

\begin{table}
	\centering
	\caption{Time lag measurements in the observed-frame}
	\begin{tabular}{llcc}
		\tableline\tableline
		 & & \lagpeak & \lagcen \\
		 \tableline
		\Hb & Year 2 & $8.52^{+2.23}_{-2.27}$ & $8.46^{+2.08}_{-2.14}$ \\ 
		\HeII & Year 2 & $0.92^{+1.58}_{-2.42}$ & $0.84^{+2.05}_{-2.19}$ \\
		\FeII & Year 2 & $3.51^{+1.49}_{-2.76}$ & $3.74^{+1.97}_{-2.62}$ \\
		\tableline
	\end{tabular}
	\label{table4}
\end{table}

\subsection{Line Width Measurements}

\begin{figure}
	\includegraphics[width = 0.48\textwidth, height = 0.55\textwidth]{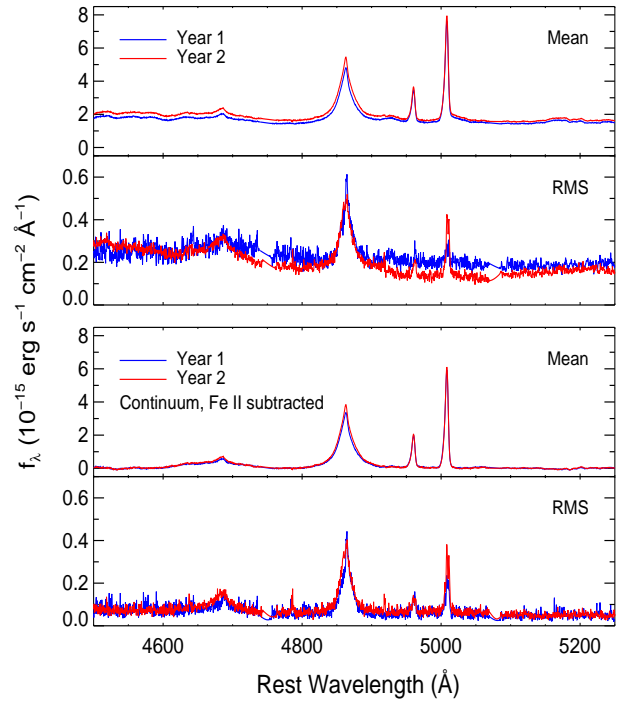}
	\caption{Mean and rms spectra of PG 0934+013, based on the individual spectra without removing power-law and \FeII\ components (top)
and constructed from individual spectra after removing power-law and \FeII\ components (bottom), using Year 1 data (blue) or Year 2 data (red), 
respectively. All individual spectra are rescaled based on the \OIII\ line.
		\label{fig5}}
\end{figure}

To measure the line widths, we constructed mean and rms spectra following the equations below, 
\begin{equation}\label{eq:mean_spec}
{\langle f(\lambda) \rangle} = \frac{1}{N}\sum_{i=1}^{N}f_{i}(\lambda), 
\end{equation}
\begin{equation}\label{eq:rms_spec}
\delta(\lambda)  = \Bigg[ \frac{1}{N-1}\sum_{i=1}^{N} \big[f_{i}(\lambda)-{\langle f(\lambda) \rangle} \big]^{2} \Bigg]^{\frac{1}{2}},
\end{equation}
where $f_{i}(\lambda)$ is the flux of $i$th spectrum and $N$ is the total number of spectra. For Year 1, we excluded the first five spectrum for constructing the mean and rms spectra. Due to the different instrumental set-up, those spectra can cause significant systemic uncertainty, particularly for the left wing of \Hb\ line.
In Figure~\ref{fig5}, we present the mean and rms spectra of PG 0934+013, respectively for Year 1 and Year 2. In the upper panels, we first show the mean and rms spectra, which are made from individual spectra without subtracting power-law continuum and \FeII\ components. The rms spectrum reflects the variability of AGN continuum and \FeII\ components as the flux level is significantly larger than zero. In contrast, the rms spectrum in the bottom panel, which is made of individual spectra after subtracting power-law 
and \FeII\ components, shows much cleaner variability of the lines, e.g., \Hb\ and \HeII. Note that \OIII\ is visible in the rms spectra, indicating that the flux rescaling was not perfect. 
Note that if we used the flux rescaled spectra based on the algorithm by \citet{Fausnaugh2016}, we also obtained a very similar rms spectrum. 

To measure the line widths, we used the mean and rms spectra, which were constructed based on individual spectra after subtracting power-law and \FeII\ components. We measured both the full width at half-maximum (FWHM) and the second moment of each emission-line profile \citep[i.e., line dispersion \sigmaline;][]{Peterson2004}, which is defined as \begin{equation}\label{eq:linedisp}
\sigma_{\rm line}^{2}  = {\frac{\sum \lambda_{i}^{2} f_{i}}{\sum f_{i}} - \lambda_{0}^{2}}  ,
\end{equation}
where $f_{i}$ is the flux density and $\lambda_{i}$ and ${\lambda}_{0}$ is the flux weighted centroid wavelength of the line profile.
To measure line dispersion, we fitted the mean and rms spectra with a power-law model using the spectral regions which represents the continuum in the vicinity of of the \Hb\ line, in order to subtract the continuum.
The uncertainty of line width measurements were determined by using the Monte Carlo bootstrap method \citep{Peterson2004}. From a set of $N$ spectra, we randomly selected $N$ spectra without considering previous selection. We then made mean and rms spectra for each realization
to measure the line width. After constructing the distribution of line width measurements from 10,000 realizations, we adopted the standard deviation of the distribution as uncertainty.
Note that the FWHM measurements using the rms spectra are relatively uncertain since the peak and the half-maximum are not straight forward to determine \citep{Peterson2004}. Thus, we provide FWHM and line dispersion measurements using the total \Hb\ profile (without removing a narrow component) and the broad component of the line profile, respectively.
Note that we constructed the rms and mean spectra, with/without removing the narrow component of \Hb\ from individual single-epoch spectra, respectively, for measuring the width of the total \Hb\ line profile or the broad component of \Hb\ (see Table 5).

We present the line width measurements without spectral resolution correction in Table 5, to avoid any systematic uncertainty of the resolution in comparing line width measurements. For black hole mass calculation, we corrected for the spectral resolution (R=1040), which reduced the line width by only a few \%.

In the case of the \sigmaline\ of \Hb, the measurements are similar regardless of the removing the narrow component of \Hb\ since the second moment does not significantly depend on the peak of the line, while the \sigmaline\ measured from the mean spectra is significantly larger than that measured from the rms spectra. Similar discrepancy was reported in the previous studies \citep[e.g.,][]{Bentz2006,Pei2014}, suggesting that there may be a difference between the varying and non-varying parts of the BLR \citep[for example, see the discussion in][]{Park+12}.

For black hole mass determination, the second moment of the \Hb\ line in the rms spectrum is commonly used as the measure of line width because it is less sensitive to peaky line profiles. In this study we use the \sigmaline\ of \Hb, measured from the rms spectrum, which was constructed
from individual spectra without removing the narrow component (bottom panels in Figure 6). Even if we remove the narrow component from \Hb\ and construct the rms spectrum, the \sigmaline\ of \Hb\ remains almost the same (see Table 5). 
In Table 6, we also present the line width measurements based on the mean and rms spectra, which are made of individual spectra after
flux-rescaling using the code by \citet{Fausnaugh2016}. Based on this comparison we conclude that line width measurements are virtually the same regardless of the flux rescaling method.

We also measured the velocity and uncertainty of \HeII\ in the same manner (see Table 7).
Since the narrow component in \HeII\ can be easily separated from the broad component in the \HeII\ line profile,
we measured the FWHM and line dispersion from the broad component of \HeII, 
which were larger than those of \Hb\ by a factor of two on average, 
indicating that \HeII\ emitting region is closer to the accretion disk than \Hb\ emitting region as expected from the stratification \citep{Wandel1997,PetersonWandel2000}.

\begin{table}
	\centering
	\caption{Line width measurements from the total \Hb\ line profile or the broad component of \Hb}
	\begin{tabular}{lccccc}
		\tableline\tableline
		&  & \multicolumn{2}{c}{$\Delta V$ (BC+NC)}  & \multicolumn{2}{c}{$\Delta V$ (BC only)}  \\
		& & FWHM & \sigmaline & FWHM & \sigmaline \\
		& & (\kms) & (\kms) & (\kms) & (\kms) \\
		\tableline
		Year 2 & mean  & 1138 $\pm$ 14 & 1126 $\pm$ 10 & 1718 $\pm$ 23 & 1168 $\pm$ 16 \\
		& rms   & 930 $\pm$ 40 & 679 $\pm$ 45 & 1358 $\pm$ 49 & 641 $\pm$ 46 \\
		Year 1+2 & mean & 1129 $\pm$ 13  & 1117 $\pm$ 8 & 1708 $\pm$ 20 & 1160 $\pm$ 12 \\
		& rms  & 1030 $\pm$ 45 & 660 $\pm$ 46 & 1449 $\pm$ 47 &  653 $\pm$ 46 \\
		\tableline
	\end{tabular}
			\tablecomments{The line widths (FWHM and $\sigma_{line}$) are measured from either mean or rms spectra, based on Year 2 data or the combined data of Year 1 and Year 2. Note that these measurements are presented without spectral resolution correction.}
	\label{table5}
\end{table}

\begin{table}
	\centering
	\caption{Line width measurements from the total \Hb\ line profile after flux-rescaling with the \citet{Fausnaugh2016} algorithm.}
	\begin{tabular}{lccc}
		\tableline\tableline
		&   & \multicolumn{2}{c}{$\Delta V$ (BC+NC)}    \\
		& & FWHM & \sigmaline\\
		& & (\kms) & (\kms)  \\
		\tableline
		Year 2 & mean  & 1171 $\pm$ 15 & 1107 $\pm$ 9 \\
		             & rms   & 959 $\pm$ 40 & 696 $\pm$ 46 \\
		Year 1+2 & mean & 1143 $\pm$ 18  & 1121 $\pm$ 10  \\
		                 & rms  & 1011 $\pm$ 44 & 659 $\pm$ 45 \\
		\tableline
	\end{tabular}
		\tablecomments{The line widths (FWHM and $\sigma_{line}$) are measured from either mean or rms spectra, which are
		made of individual spectra after flux-rescaled with the code by \citet{Fausnaugh2016}. Note that these measurements are presented without spectral resolution correction.}
	\label{table5_f}
\end{table}

\begin{table}
	\centering
	\caption{\HeII\ line width measurements}
	\begin{tabular}{lccc}
		\tableline\tableline
		&  Spectrum  & \multicolumn{2}{c}{$\Delta V$ (BC only)}  \\
		& & FWHM & \sigmaline \\
		& & (\kms) & (\kms) \\
		\tableline
		Year 2 & mean & 5093 $\pm$ 31 & 2116 $\pm$ 23 \\
		& rms   & 2484 $\pm$ 53 & 1030 $\pm$ 47 \\
		Year 1+2  & mean &  5683 $\pm$ 23 & 2031 $\pm$ 34 \\
		& rms  & 2487 $\pm$ 46 &  1045 $\pm$ 45 \\
		\tableline
	\end{tabular}
	\tablecomments{Note that these measurements are presented without spectral resolution correction.}
	\label{table6}
\end{table}

\subsection{Black Hole Mass Determination}

Assuming virial theorem, the mass of central black hole can be determined as
\begin{equation}\label{eq:mbh}
M_{\rm BH}=f\frac{c \tau ({\Delta V})^2}{G}\ 
\end{equation}
where f is a virial factor, which depends on the structure and kinematics of the BLR, $\Delta V$ is the width of broad emission line, $\tau$ is the time lag between continuum and emission line, c is the speed of light, and G is the gravitational constant. 
In order to determine \mbh, we employed the rest frame \lagcen\ of \Hb, after converting the measured time delay to the rest-frame time lag. We also used the line width of \Hb\ (i.e., \sigmaline\ and FWHM, respectively) in the rms and mean spectra, which were generated from single-epoch spectra with/without removing the narrow component of \Hb\ (see Table 5), after correcting for the spectral resolution. Here, we adopt the virial factor, f = 4.47 for the line dispersion of \Hb\ and f = 1.12 for the FWHM of \Hb\ based on the calibration by \citet{Woo2015}. These virial factors  were derived by normalizing the virial products to the \mbh\ - \sigmastar\ relation of local galaxies with dynamical black hole mass measurements \citep[e.g.,][]{Onken2004,Woo2010,Woo2015,Park2012,Grier2013}. The uncertainties of \mbh\ were determined via error propagation from the uncertainty of the time lag and line width measurement. 

The derived black hole mass ranges from $\sim 1.4 \times 10^{6}$ to $\sim 9.5 \times 10^{6}$ \msun\ depending on the choice of the velocity measurements (see Table~8). As the best measurement, we combined \sigmaline\ measured from the total \Hb\ line profile using the rms spectrum and the \lagcen\ measurement from Year 2 data, which provides \mbh\ = $3.13^{+0.91}_{-0.93} \times 10^{6}$\msun. On the other hand, if we use the FWHM measurement from the broad \Hb\ component based on the rms spectrum, we obtained \mbh\ =  $3.10^{+0.83}_{-0.85} \times 10^{6}$\msun. These two masses are consistent within the 1$\sigma$ error. If the line width measurements from the mean spectrum are used, we obtain somewhat higher black hole masses due to the higher values of the line widths \citep[see also][]{Sergeev+99, Shapovalova+04, Collin+06, Park+12}.

\begin{table}
	\centering
	\caption{Black hole mass determination based on Year 2 data}
	\begin{tabular}{lccc}
		\tableline\tableline
		Spectrum  & $\Delta V$ & \mbh\ (BC+NC)  & \mbh\ (BC only)  \\
		& & (\msun) & (\msun) \\
		\tableline
		Mean & FWHM &  $2.13^{+0.55}_{-0.57} \times 10^{6}$ & $5.05^{+1.31}_{-1.35} \times 10^{6}$  \\
              & \sigmaline & $8.80^{+2.28}_{-2.34} \times 10^{6}$ & $9.47^{+2.46}_{-2.53} \times 10^{6}$ \\
           rms & FWHM & $1.38^{+0.38}_{-0.39} \times 10^{6}$ & {\bf $3.10^{+0.83}_{-0.85} \times 10^{6}$  } \\
              & \sigmaline & {\bf $3.13^{+0.91}_{-0.93} \times 10^{6}$ } &  $2.78^{+0.82}_{-0.84} \times 10^{6}$  \\
		\tableline
	\end{tabular}
	\label{table8}
\end{table}

\subsection{Velocity-resolved Lag}
We investigate whether we can constrain the kinematic structure of the BLR by measuring the \Hb\ velocity-resolved lags. 
After dividing the \Hb\ line profile in the rms spectrum into several velocity channels, of which the velocity ranges were determined to have the same flux in each channel in the rms spectrum (see Fig. 7), 
we constructed a light curve for each velocity channel to cross-correlate it with the continuum light curve and obtained 
the time lag and uncertainty for each velocity channel using the same Monte Carlo realizations as we used in S 3.2.
We experimented with the number of the velocity channels, e.g., seven, five, and three, in order to reduce the uncertainty of the time lag measurements.
However, the uncertainty of the time lags from each velocity channel is significantly larger than the difference of the time lags among velocity channels.
As an example, we present the velocity resolved time lag measurements using three velocity channels, reporting that we obtained
no meaningful trend due to the large error bars. 

As a consistency check, we used the rms spectrum, which is made of individual single-epoch spectra after removing the narrow component of 
\Hb\ (blue line in Figure 7), in order to investigate the effect of the narrow component. The rms spectrum without the narrow \Hb\ component is less peaky than that with the narrow \Hb\ component. We interpret that this is due to the uncertainty of removing the narrow component in each single-epoch spectrum.
 The narrow component can be subtracted somewhat more or somewhat less from each spectrum, and the fluctuation of this residual
 component from the narrow \Hb\ compensates the variability of the broad \Hb\ component. 
 As we discussed in Section 3.2, this is why we used the spectra without removing the narrow \Hb\ component for constructing the \Hb\ light curve and measuring the time lag of \Hb. Nevertheless, we obtained consistent lag measurements within the error, regardless of removing the narrow \Hb\ component. 

\begin{figure}
	\includegraphics[width = 0.48\textwidth]{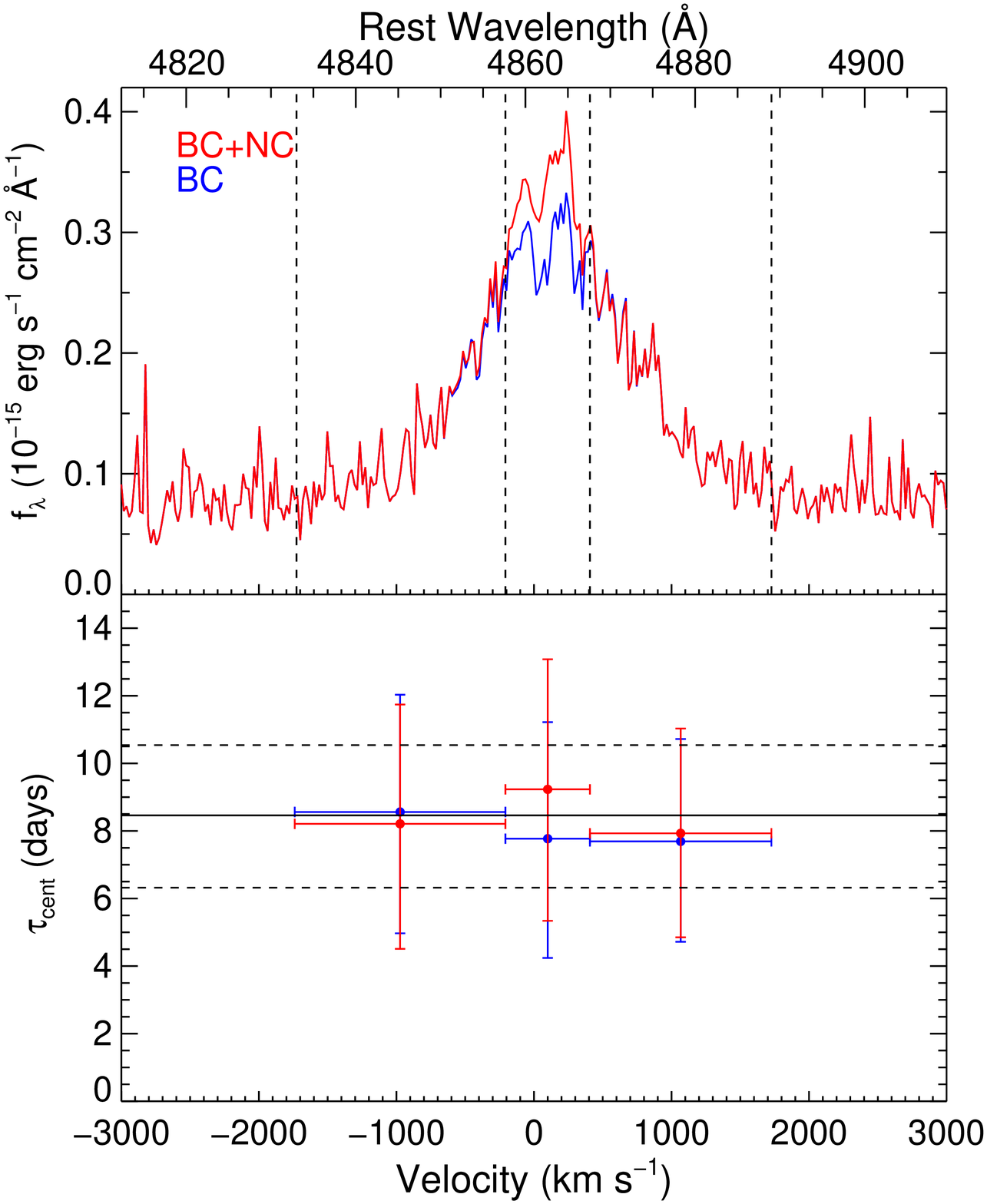}
	\caption{Velocity-resolved reverberation in the \Hb\ line. The upper panel shows the \Hb\ line profile in the rms spectrum without removing the narrow component (BC+NC; red line). In comparison we also present the rms spectrum made of individual spectra after removing the narrow \Hb\ component (BC: blue line). Three velocity channels are denoted with dashed lines. The lower panel shows  the time lag measured for each velocity channel with the horizontal error bar representing the width of the each velocity bin. The time lag based on the total line profile of \Hb\ and its uncertainty are marked with solid and dashed lines,
	respectively.  
		\label{fig6}}
\end{figure}

\section{Discussion and conclusions}

\subsection{Mass determination}

We performed photometric and spectroscopic monitoring observations over two sets of 4 month campaign to measure the emission line time lags and determine black hole mass of PG 0934+013. 
Based on the interpolation CCF analysis, we measured the centroid lag of \Hb\ as $8.06^{+1.98}_{-2.04}$ days in the rest-frame based on Year 2 data, while the time lag from Year 1 data was not reliably determined. Using the total line profile of \Hb\ in the rms spectrum,
we obtained \sigmaline\ = 668$\pm$44 \kms\ after correcting for the spectral resolution.
By applying the virial factor f = 4.47 from the recent calibration based on
the \msigma\ relation by \citet{Woo2015}, we determined \mbh\ = $3.13^{+0.91}_{-0.93} \times 10^{6}$\msun.

The \mbh\ of PG 0934+013 was previously estimated with single-epoch method in the literature. For example, \cite{Vestergaard2006} derived \mbh\ = $7.041^{+0.092}_{-0.117} \times 10^{7}$\msun, while \cite{Veroncetty2001} reported \mbh\ = $4.3 \times 10^{6}$\msun.
If we use the SDSS spectrum of PG 0934+013, we derive \mbh\ = $1.3 \times 10^{7}$\msun\ using \fwhb\ and \lam\lf. Note that the single epoch \mbh\ estimate varies due to a number of sources of systematic difference, i.e, a different virial factor \citep[see][]{Park2012}, various methods of measuring the H$\beta$ line width, and the different calibration of the size-luminosity relation \citep[see][]{Bentz2013}. Nevertheless, these single-epoch mass estimates are roughly consistent with our \mbh\ measurement based on the reverberation mapping result within a factor of several.

\subsection{Size-luminosity relation}

We investigate whether the measured \Hb\ lag is consistent with the size-luminosity relation.
Using the measurement of $\lambda$\lf\ = $4.26 \times 10^{43}$ \ergs\ from the mean spectrum of Year 2 data, we calculate the expected lag as $22.6^{+0.7}_{-0.7}$ days, based on the \Hb\ lag-luminosity relation of \citet{Bentz2013}. 
The expected lag is larger than our measurement of $8.46^{+2.08}_{-2.14}$  days by a factor of a few.
Accounting for the uncertainty of the spectroscopic flux calibration, we re-measure the $\lambda$\lf\ after re-scaling the mean spectrum based on the photometry. Since the spectral range of our data does not cover the entire \textit{B} band, we instead determine \textit{V} band magnitude from the mean spectrum and converted it to \textit{B} band magnitude with the measured color \textit{B}-\textit{V} = 0.35 from the SDSS spectrum of PG 0934+013, assuming that the color has not changed.
In this process, we determine a scale factor of 0.9 based on the mean \textit{B} band magnitude of 16.40 $\pm$ 0.07  from our photometric campaign in Year 2. If we use the scaled $\lambda$\lf, then the expected lag becomes $21.85^{+0.68}_{-0.67}$ days. 
 
\begin{figure}
	\includegraphics[width = 0.45\textwidth]{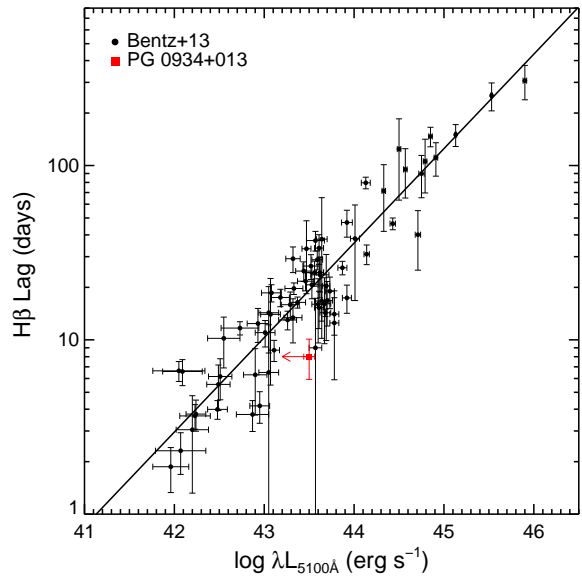}
	\caption{Size-luminosity relation adopted from \citet{Bentz2013} along with our measurements of PG 0934+013 (red square). We adopted the AGN luminosity estimated based on the HST WFC3 F438W image. Since the host galaxy contribution is relatively uncertain, we denoted the luminosity 
	with an upper limit sign. 
		\label{fig7}}
\end{figure}

In Figure~\ref{fig7} we present the size-luminosity relation of the reverberation-mapped AGNs from \citet{Bentz2013} along with our measurements of PG 0934+013. We point out that the $\lambda$\lf\ measurement is relatively uncertain. The source of uncertainty of the $\lambda$\lf\ includes the spectroscopic flux calibration and the rescaling of the flux based on photometry, and the variability of the source during the campaign. The rms variability of the source during the campaign is larger than the typical measurement error of the flux at 5100\AA\ in each spectrum.
Note that if the host galaxy stellar continuum is not subtracted from the observed flux, the measured $\lambda$\lf\ can be considered as an upper limit. For luminous AGNs, stellar contribution to $\lambda$\lf\ is negligible. However, it is necessary to correct for the host galaxy contribution for moderate to low luminosity AGNs such as PG 0934+013 \citep[e.g.,][]{Bentz2013}. Although we included a stellar population model in the spectral decomposition process for several epochs as these spectra with a high S/N ratio showed stellar absorption features, we were not able to reliably quantify stellar fraction in the monochromatic luminosity at 5100\AA\ based on these spectra.

\begin{figure}
	\includegraphics[width = 0.49\textwidth]{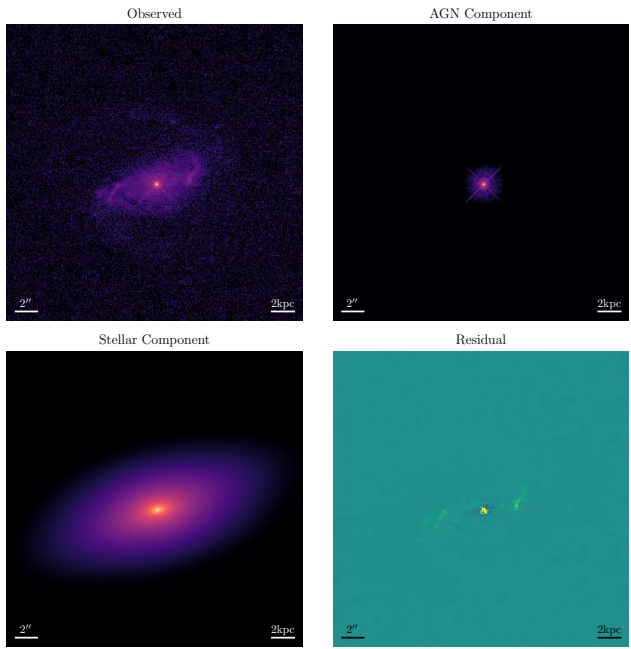}
	\caption{2-D decomposition of PG 0934+013 (upper left), using a PSF for AGN (upper right) and two components (bulge and bar) for the host galaxy (lower left)
	based on the F438W band image obtained with the HST WCF3. 
}
\end{figure} 

Deep high resolution imaging data are required to provide a robust measurement of $\lambda$\lf\ from AGN component, via, for example, the 2-D surface bright fitting analysis. The available HST WFC3 archival images of PG 0934+013 are relatively shallow with a 450 second exposure in the F438W band and a 300 second exposure in the F814W band, which are not sufficient to obtain rigorous results with the AGN-galaxy decomposition analysis based on our experience \citep{Bennert2010,Park2015}. Nevertheless, we tried to use both F438W and F814W band images of PG 0934+013 to estimate the AGN fraction. Since a stellar bar is clearly present in the raw image, we used two components for the host galaxy, namely, a bulge (Sersic index n=4) and a bar (n=0.5), along with a PSF component, which was constructed with the TinyTim, for a 2-D surface brightness fitting with the GALFIT \citep{Peng02}. Although it is difficult to constrain the host galaxy structure, we were able to obtain a reasonable fit with the three components albeit with a large residual, particularly at the spiral structure, and measured the fraction of the PSF component as presented in Figure 9.
The AGN fraction is 0.38 in the total flux observed through the F438W band. In the case of the F814W band image, the AGN fraction is 0.42. If we apply the same aperture used for spectroscopy, this fraction increases to 0.71 and 0.61, respectively for F438W and F814W images since the host galaxy flux is more widely distributed than the point source. 
On the other hand, if we convolve with a typical ground-based seeing, the AGN fraction may also change depending on the exact radial profile of the host galaxy.  We experimented with a series of seeing size to degrade the HST image to investigate the seeing effect. We find that the AGN fraction
changes insignificantly within a several $\%$ level. 
Although the AGN fraction is wavelength-dependent, if we take the face value of the AGN fraction (i.e., 0.42) based on the F814W band image in order to correct for the host galaxy contribution, then the AGN continuum luminosity becomes $\lambda$\lf\ = $1.61 \times 10^{43}$ \ergs, and the expected lag from the size-luminosity relation is $\sim$13 days. Based on the F438W band magnitude of the AGN from the decomposition analysis, we determine the continuum luminosity at 5100\AA, using the SDSS spectra after rescaling the flux to match the F438W band magnitude. The obtained luminosity is $\lambda$\lf\  = $3.14 \times 10^{43}$ \ergs, as shown in Figure 8. 
Considering the flux uncertainties and systematic effects, i.e., host galaxy contribution, we find no strong evidence that PG 0934+013 deviates from the size-luminosity relation.
A more detailed analysis on the host galaxy
based on a high quality imaging data is necessary to confirm whether PG 0934+013 follows the size-luminosity relation.

Recent reverberation studies on narrow-line Seyfert 1 galaxies claimed that super-Eddington AGNs have shorter \Hb\ lags than the expected lag from the size-luminosity relation \citep{Du2015,Du2016}. However, since the Eddington ratio of PG 0934+013 is $\sim$1\%, which is calculated using the observed luminosity at 5100\AA\ ($\lambda$\lf) without considering host galaxy contribution after multiplying a bolometric correction factor 10 \citep{WooUrry2002}, it is not relevant to compare the target with the high Eddington AGNs in the context of the size-luminosity relation.

\subsection{\msigma\ relation}

\begin{figure}
	\includegraphics[width = 0.45\textwidth]{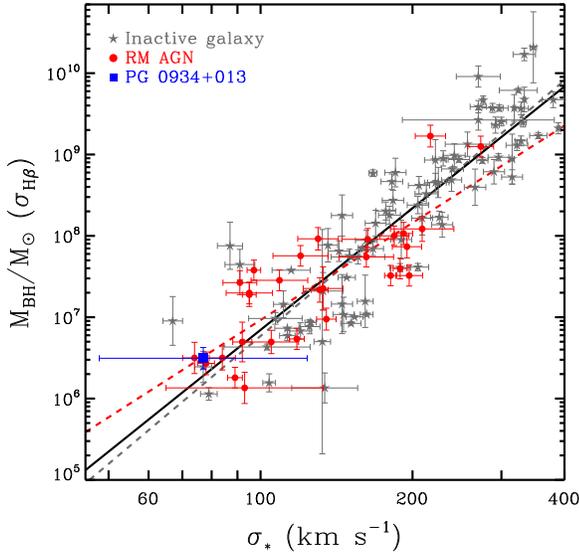}
	\caption{\msigma\ relation of reverberation-mapped AGNs (red filled circles) and inactive galaxies (grey stars)
	adopted from  \citet{Woo2015}, including our measurement of PG 0934+013 (blue square). 
	The solid line is the best fit from \citet{Woo2015}, while the dashed lines represent the best fit respectively for 
	reverberation-mapped AGNs (red) and inactive galaxies. 
		\label{fig8}}
\end{figure}

Finally, we discuss whether PG 0934+013 follows the scaling relations between \mbh\ and galaxy properties. Since the stellar 
velocity dispersion of PG 0934+013 has not been reported in the literature, we tried to measure it using our own spectrum obtained with the best sky condition although the stellar continuum is still very noisy. After masking out the emission line [N I] \lam5201, we obtained $\sigma_{*} = 77\pm34$ \kms\ from the stellar absorption lines in the spectral range 5150-5300 \AA, using the penalized pixel-fitting (pPXF) method and the MILES stellar population models with a solar metallicity \citep{CappellariEmsellem2004}.
In Figure~\ref{fig8}, we present the  \mbh-stellar velocity dispersion relation defined with a joint-analysis of inactive and active galaxies from \citet[][see also \citet{Woo+13}]{Woo2015}, along with our measurements of PG 0934+013. While the stellar velocity dispersion measurement is highly uncertain due to the low S/N on stellar continuum and relatively low spectral resolution (i.e., line dispersion resolution $\sigma$=122 \kms), we find no significant evidence that whether PG 0934+013 offsets from the \msigma\ relation defined by inactive galaxies and other reverberation-mapped AGNs \citep{Woo+13,Woo2015}. Further investigations with much better data and measurement is necessary to investigate the case.

~

Our reverberation study demonstrates that large telescopes equipped with a queue observation mode 
can be effectively used for time series studies of AGNs. The capability of the SALT for regular access
and sensitivity will be very useful for future reverberation studies of both low luminosity AGNs, which requires a short cadence, and for high-z AGNs, which requires a large photon-collecting power.

\acknowledgments
We thank the anonymous referee for various suggestions, which were helpful to improve the clarity of the paper.
This work was supported by Samsung  Science and Technology Foundation under Project Number SSTF -BA1501-05
and the National Research Foundation of Korea grant funded by the Korea government (No. 2016R1A2B3011457).
E.R.C. and S.C. gratefully acknowledge the receipt of research grants from the National Research Foundation (NRF) of South Africa.
C.C. and M.I. acknowledges the support from National Research Foundation of Korea grant, No. 2008-0060544, funded by the Korean government (MSIP). This work is partially based on the data collected from the Mt. Lemmon Optical Astronomy Observatory operated by Korea Astronomy and Space Science Institute. Spectroscopic observations reported in this paper were obtained with the Southern African Large Telescope (SALT).


\end{document}